# Measuring salinity and density of seawater samples with different salt compositions and suspended materials.

Aleksandr N. Grekov [1,*], Nikolay A. Grekov [1] and Evgeniy N. Sychov [1]

Institute of Natural and Technical Systems; oceanmhi@ya.ru

* Correspondence: oceanmhi@ya.ru; Tel.: +79787883197

*Determining the weight of dissolved materials in seas and oceans through in-situ measurements remains an unsolved problem. To solve it, it is necessary to develop new methods and metering devices. This article analyzes the indirect salinity and density measurement methods using measured in-situ parameters (relative electrical conductivity, sonic velocity, temperature, and hydrostatic pressure). The authors review an electric conductivity sensor design that allows for obtaining data on suspended irregularities along with measuring the impedance of electrodes under various alternating current frequencies. The authors analyze the joint metering technique using the CTD and SVP devices in a marine testing area. Following the results of joint metering, the authors present a test of water samples with different salt compositions for the presence of suspended matter.*

During the field research in the Black Sea and the Sea of Azov, we conducted an experiment to determine the reliability of the measurement results obtained using the CTD and SVP devices in the shelf zone. We used these data to calculate the salinity and density of seawater.

When studying the results of joint metering using the CTD and SVP devices, we identified a number of factors that signified an incorrect interpretation of the data obtained. We compared the values of salinity and density obtained indirectly through well-established algorithms.

We made an assumption that the use of standard algorithms that do not consider the effects of various suspended materials in the seawater results in significant errors in the calculations of salinity and density when processing *in situ* measurement results obtained using the CTD and SVP devices.

We must note that some researchers have been trying to develop methods to correct field-measurement results that would help determine the true values of absolute salinity and density of seawater with unconventional composition, including samples with suspended matters.

Some well-known works [1-9] solve this problem only partially. For example, when calculating salinity anomaly $\delta S_A$ with CTD probes with electric conductivity measurement channels, the user manual for TEOS-10 [15] suggests using a chemical model of conductivity and density [1] to assess the correlation between salinity changes and measured properties of seawater.

This model allows for determining the salinity anomaly expressed in [*g/kg*] through nitrate ($NO_3^-$) and silicate ($Si(OH)_4$) concentration values in the seawater sample, as well as the difference between $\Delta TA$ and $\Delta DIC$ expressed in [*mole/kg*]:

$$\delta S_A / (g/kg) = (\, 55.6\, \Delta TA + 4.7\, \Delta DIC + 38.9\, NO_3^- + 50.7\, Si(OH)_4\, ) / (mole/kg) \qquad (1)$$

In this formula, $\Delta TA = TA - 0.0023\,(S_P/35)$, *mole/kg*; $\Delta DIC = DIC - 0.00208\,(S_P/35)$, *mole/kg*; where $S_P$ is the practical salinity expressed in practical salinity units [*psu*]. Differences $\Delta TA$ and $\Delta DIC$ are determined between the Total Alkalinity (TA) and Dissolved Inorganic Carbon (DIC) in the sample and the respective best assessments of TA and DIC in standard seawater [1, 2]:

According to [2], the standard uncertainty of model compliance is 0.08 mg/kg in the oceanic range if the exact amounts of all the substances are known.

Other research works showed that $\delta S_A$ can be calculated using a simplified empirical equation based on the measurement of $SiO_2$ silicate concentration measurement and the total alkalinity TA because these components are measured and well-suited for the assessment of density and salinity changes of the deep water.

This empirical equation looks as follows [3]:

$$\delta S_A = a\,\Delta[\text{SiO}_2] + b\,\Delta[\text{NTA}], \tag{2}$$

where NTA is the total alkalinity, normalized (modified) to salinity 35 (NTA=(TA / Sp) 35).

According to Millero [4–7], $\Delta[\text{NTA}]$ is the difference between the measured value of normalized total alkalinity (NTA) and the reference value (2300 *umole/kg*) for surface seawater. However, the practical method of determining $\delta S_A$ is based on the use of density measurements [8].

As stated in [3, 9] the method proposed by Millero *et al.* can be used to calculate $\delta S_A$ through the following formula: $\delta\rho = 0.75179\,\delta S_A$.

Unfortunately, the majority of the results were obtained in laboratories because it is hardly possible to adjust the measurement results through bathometer sampling and subsequent seawater density measurements in the field, onboard a ship, using, for instance, Anton Paar DMA 5000 M vibration sensor because keeping the impurities in the sample suspended is difficult.

It is difficult to use these models and research works when working *in situ* in the shelf zone where there are suspended impurities.

As mentioned above, our research was conducted in the natural conditions of the shelf zone using two devices and one model.

To do this, we complemented a standard SBE911 oceanological probing CTD unit [10] with an ISZ-1 SVP probe developed at the Institute of Natural and Technical Systems (INTS) [11] (Figure 1). The ISZ-1 probe also featured an ultrasonic attenuation duct. Unfortunately, it only worked on a few of the stations and it ran in the testing mode. The data on the metering attenuation duct are provided in [12]. In the testing mode, an *impedance* channel model with the 4-electrode mesh was used, and its operation is described below.

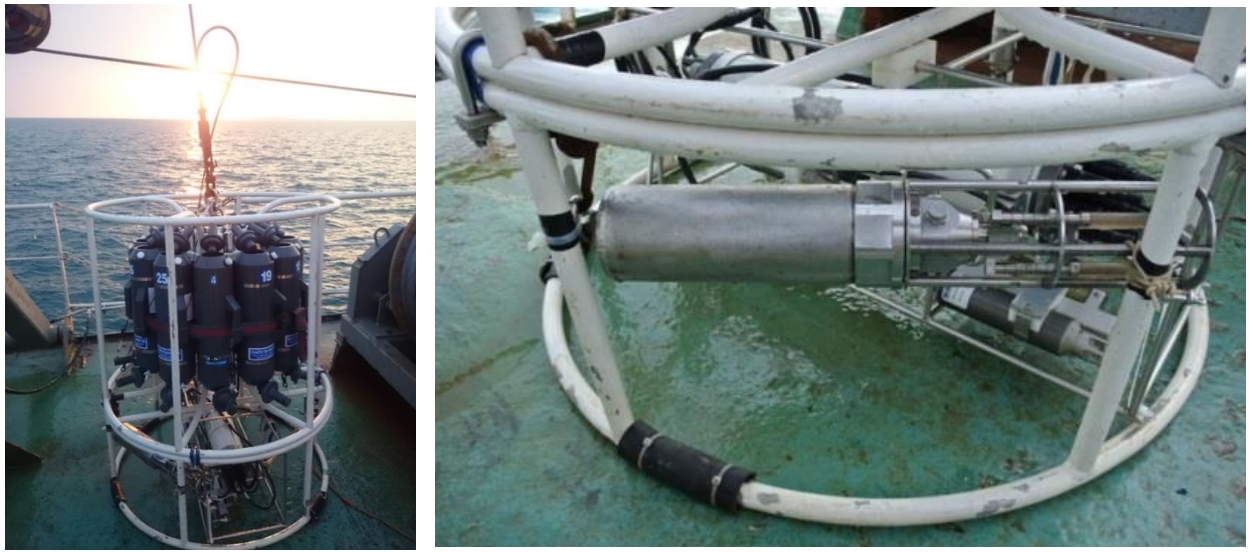

Figure 1. The general view of the joint mount for the SBE911plus probing CTD unit and the attached ISZ-1 SVP probe developed at the INTS

The specifications for both of the probes are presented in Tables 1 and 2.

Table 1. Specifications of ISZ-1 probe

| Measured and calculated parameters | Range measured | Random error | Error |
|---|---|---|---|
| Sonic speed, m/s | 1375 ÷ 1900 | 0.001 | ± 0.02 |
| Water temperature, °C | -2 ÷ +35 | 0.001 | ± 0.01 |
| Hydrostatic pressure, dbar | 0 ÷ 2000 | 0.2 | ± 2 |

Table 2. Specifications of SBE911plus

| Measured parameters | Range measured | Error |
|---|---|---|
| Electric conductivity, mho/m | 0 – 7 | 0.0003 |
| Water temperature, °C | -5 ÷ +35 | 0.001 |
| Hydrostatic pressure, dbar | 0 ÷ 2000 | 0.015% |

Both devices were mounted on the same basket next to one another, which allowed us to compare measurement results based on two different salinity and density calculation methods using electric conductivity or sonic speed. When calculating density and salinity using the data measured by the CTD probe, we used the integrated SeaBird software based on TEOS-10.

When reviewing measurement results obtained using the CTD and SVP devices to determine the causes of discrepancies, we also analyzed the measurement channels for electrical conductivity and sonic speed in the water. Two other measurement channels (pressure and temperature) were identical in their metrological parameters. To compare the discrepancies in the measurements obtained using the CTD and SVP devices, we compared the calculated values of salinity and density using the algorithms based on TEOS-10 and well-tried for clear seawater without impurities. Let us review in detail what factors affect the reliability of measurements obtained through *in situ* devices by analyzing the operations of measurement channels for sonic speed and electrical conductivity.

CTD probes widely employ the contact (conductive) method to measure the specific electrical conductivity of water, as well as the contactless (inductive) method. For example, conductive sensors are used in devices like Neil Brown Instrument Systems (NBIS), Sea Bird Electronics Inc (SBE), Guildline Instruments Ltd (Guildline), IDRONAUT S.r.l, etc. Inductive sensors are used in Aanderaa Data Instruments AS, Falmouth Scientific, Inc, and other devices. Despite their advantages (zero contact), inductive sensors have a narrow frequency bandpass, and it is very difficult to extend their functional capacities, e.g. use them for impedance spectroscopy.

The most successful conductive meshes have four electrodes or more. Electrodes are made of platinum and located inside an alumina ceramic or borosilicate glass cylindrical tube. For instance, the mesh of Sea Bird Electronics Inc (SBE) features circular platinized electrodes that are 10 mm wide, with a tube length of 190 mm and an internal diameter of 7 mm.

Since the meshes used have almost zero external field impacts, the sensor has good metrological parameters. However, the large mesh length and a small cross-section of the flow channel make it impossible for water to pass through it with the required speed. Thus, forced injection systems (special pumps) are used in this case. The small diameter of the flow channel makes SBE probes very sensitive to impurities.

When electric current passes through a conductive mesh, a chain of electric and thermodynamic processes starts in this mesh.

While the current is not supplied to the electrodes, ions in the solution only take part in thermal motion. When the mesh is supplied with a field with intensity E, it makes the ions move to the electrodes, complementing the thermal motion.

The key problem with conductivity measurement using alternating current is the correct interpretation of the results. It can be complicated by the fact that the equivalent circuit of a mesh is usually unknown and that the sample with connected electrodes is basically an electric black box.

When studying the impedance of electrochemical meshes, it is necessary to obtain reliable information on electrode processes, i.e. the processes occurring on the surface of contact between the electrode and electrolyte.

Let us review in detail the conductive sensor with 4-electrode mesh.

The equivalent circuit of a 4-electrode mesh with an auxiliary node to measure impedance is shown in Figure 2.

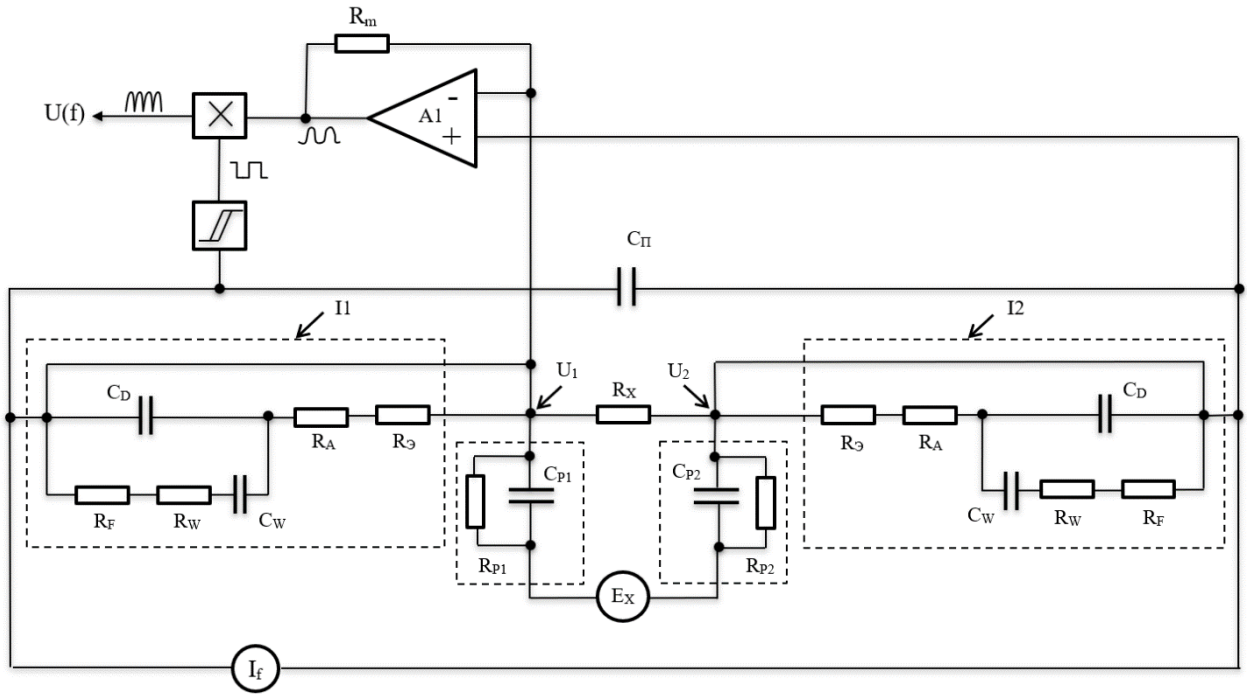

Figure 2. The equivalent circuit of a 4-electrode mesh with an auxiliary node to measure impedance

The equivalent circuit of the mesh in question comprises two current (I1 and I2) and potential (U1, U2) electrodes with some of the impedance of the water solution in question between them. Current electrodes I1 and I2 are supplied with alternating current. They can be represented as a capacitance of double layer $C_D$, parallel with the impedance of electrochemical polarization $R_F = \frac{\partial E}{\partial i_F}$ and serial-connected to resistance $R_W$ and capacitance $C_W$ related to the concentration polarization (usually referred to as Warburg's impedance) [13].

Resistance $R_A$ reflects the absorption of atoms, ions, or molecules on the surface of the electrode, and it can be equal to zero in some cases.

In an equivalent circuit, parasitic capacitance $C_{par}$ accounts for a sum of capacitances: the capacitance determined by the dielectric constant of the solution, the distance between the electrodes, the active surface areas of electrodes, and the capacitance between the conductors connected to the electrodes. When considering impedance, it is worth noticing that the parasitic capacitance is especially evident at higher frequencies. For the equivalent circuit shown in Figure 2, the total impedance of the conductometric mesh in question can be expressed as follows, provided that $C_{par} = 0$:

$$Z_я = R_э + \frac{2R_W C_W^2}{\omega^2 R_W^2 C_W^2 C_D^2 + (C_D + C_W)^2} - 2j\frac{\omega^2 R_W^2 C_D C_W^2 + C_D + C_W}{\omega^2 R_W^2 C_W^2 C_D^2 + (C_D + C_W)^2}. \tag{3}$$

In this expression, the first component is the true impedance value of the solution, the second real component can be denoted as $\Delta R_W$ is the error caused by polarization phenomena, and the third component is the virtual one. It can be determined as capacitive reactance.

During the metering process, capacitive reactance leads to a phase difference between the current and the voltage. It can be mitigated using special circuits introduced in the metering device and used to obtain extra information on the phase at various frequencies, which will be further used in the research of mesh impedance changeability.

We know that the material and surface condition of electrodes have a significant impact on the polarization impedance value. We assume that the value of polarization impedance is related to the design of the crystalline grid of the electrode material, the absorption properties of its active surface, and the formation of surface oxidic films. Besides, the frequency of alternating current has a significant impact on the polarization effect. Numerous authors claim that the correlation between $R_S$ and frequency for reversible electrodes made of different materials used in water solutions of various concentrations can be expressed as follows:

$$R_S = \frac{\eta}{\sqrt{\omega}},, \qquad (4)$$

where $\eta$ is a constant. From formula (4), we can conclude that $R_S$ decreases as the frequency increases and becomes insignificant at frequencies above 1 kHz. On the other hand, the correlation between the polarization capacitance and the frequency looks as follows

$$C_S = \frac{1}{\eta\sqrt{\omega}}.. \qquad (5)$$

Constant $\eta$ in expressions (4) and (5) reflects the correlation between $R_S$ and $C_S$ and the concentration, ion diffusion factor, and double-layer potential.

$$\eta = \sum_i \beta_i v_i \frac{1}{\sqrt{2D_i}}, \qquad (6)$$

where $\beta_i = \frac{\partial E}{\partial W_i}$; $W_i$ the concentration of potential determining ions of variety $i$ with a sufficient amount of inert electrolyte that is introduced to eliminate the migration effect of potential determining ions; $v_i$ is the number of ion equivalents occurring from the chemical interaction when transferring through the dividing surface of one faraday of electricity; $Di$ is the potential-determining ion concentration factor, and *E is the double layer potential*, taking into account the voltage drop ($E = \varphi = \Delta\varphi$).

Unfortunately, the linear equations provided cannot produce acceptable results when used for precise calculations in the model. Therefore, we have to settle for factors obtained in experiments for each specific case.

Relevant conditions are sometimes used to eliminate or reduce errors $R_A$ when metering ohmic resistance of mesh $R_x$ (e.g. the use of an ideally polarized electrode or inert electrolyte). As mentioned above, the polarization resistance value $\Delta R_W$, as well as the respective error, accounted in the measured resistance, depend on a large number of various system parameters.

In some cases, the error caused by the impact of polarization resistance $\Delta R_W$ on the measured resistance $R_x$ may reach 20%. Thus, for high precision measurements, it is necessary to introduce an allowance for polarization resistance $\Delta R_W$. Experiments confirmed that the minimum error caused by $\Delta R_W$ for platinum group electrodes at some frequencies is 0.001%.

Concerning design features, many authors [14] conducted experiments to prove that the distance between the electrodes in a conductometric mesh does not affect the polarization resistance value $\Delta R_W$.

We used the recommendations provided above to develop a simplified model of the electrical conductivity sensor with four platinum electrodes placed in a ceramized glass tube with an internal diameter of 8 mm and a length of 60 mm, then placed in a sealed case and tested under high pressure. This sensor design is not perfect, like, for instance, the SBE911 model described above that features a water pump. The sensor model was also equipped with an impedance measurement module operating in situ at constant frequencies within the range of 0.01–1.0 MHz. The sensor model was tested during the field trip together with the CTD and ISZ devices. It produced additional data about suspended impurities in the seawater. The impedance sensor model helped us record suspended impurities at a frequency of about 0.8 MHz, corresponding to the impurities recorded by the ultrasonic ISZ device and missing in the electric conductivity channel of the CTD probe.

The simplified function diagram of the impedance measurement module together with the equivalent 4-electrode conductometric mesh is shown in Figure 2. The impedance measurement module consists of an amplifier with high input resistance $A1$, voltage comparator and multiplier, those output sends the signal amplitude at a specific frequency characterizing the amounts of suspended matter in the seawater to the recorder. The impedance measurement module operates at 5 constant frequencies: 0.01; 0.2; 0.5; 0.8; 1.0 MHz with a homogenization duration of 1 second an each of them. The entire measurement cycle lasted for 5 seconds.

If we extend the functional capabilities of metering channels for electrical conductivity and sonic speed of the CTD and ISZ devices by using impedance and acoustic attenuation, we can obtain quantitative and qualitative parameters of suspended materials in seawater without using any additional tools.

The key measured parameters related to absolute salinity $S_A$ and density ($\rho$) include relative electrical conductance ($\chi$), sonic speed in water ($c$), temperature ($T$), and hydrostatic pressure ($P$). To calculate absolute salinity ($S_A$) and density ($\rho$) using the parameters measured *in situ*, we can use the following algorithms:

1) $S_{A1} = S_{A\chi}(T, \chi, P)$;
2) $S_{A2} = S_{Ac}(T, c, P)$;
3) $S_{A3} = S_{A\rho}(T, \rho, P)$.
4) $\rho_1 = \rho_c(T, c, P)$;
5) $\rho_2 = \rho_{SA1}(T, S_{A1}, P)$;
6) $\rho_3 = \rho_{SA2}(T, S_{A2}, P)$.

The connection between absolute salinity and practical salinity of seawater is set by the following proportion in TEOS-10 Manual [15]:

$$S_A = (35,16504 / 35)S_P + \delta S_A(x, y, P), \, g/kg. \tag{7}$$

Here $\delta S_A(x,y,P)$, g/kg is the "absolute salinity anomaly" that has to take into account the shifts in the constant salt composition of seawater. Thus, the problems associated with determining the reliable practical salinity values for seawater $S_P$, calculated taking into account conductivity and characterizing the impact of the ionic component of the dissolved materials if they pass through the 0.2 um filter is complemented by the problems of zoning the shifts in the constant salt composition of seawater and the assessment of respective absolute salinity anomaly values ($\delta S_A(x,y,P)$) aligned with geographic coordinates ($x,y$) and depth (or pressure, $P$) taking into account different suspended materials.

To calculate the density using the measurements of the ISZ device, we used the equations developed by the authors of [16].

To construct the following equation:

$$\rho = \varphi_\rho(T, P, c) \tag{8}$$

The authors used the TEOS-10 international system [15]. Based on the calculations using density and sonic speed equations from the TEOS-10 system [15], an initial data array of $\{\rho_m, c_m, T_m, P_m, S_{Am}\}_M$ was generated where M is the number of points which is over 130,000. To do this, we calculated M values for the couples of densities $\rho(T_m, P_m, S_{Am})$ and sonic speeds $c(T_m, P_m, S_{Am})$ at an interval of 1°C within the temperature range from the melting curve to 40°C, at an interval of 2 *MPa within the hydrostatic pressure range of* 0–120 *MPa* and an interval of 1 *g/kg* within the salinity range of 0–42 *g/kg*. The variation range for sonic speed was about 1300–1800 *m/s*, and for density, it was 990–1090 *kg/m³*. In the set of five parameters ($\rho, c, T, P, S_A$), only three can be used independently in any combination.

The seawater density interpolation equation suggested by us (8) was expressed as follows:

$$\gamma = \sum_{i=1}\sum_{j=1}\sum_{k=1} b_{ijk}\tau^i \pi^j \omega^k, \tag{9}$$

where $\gamma = (\rho - \rho_0)/\rho^*$; $\tau = (T - T_0)/T^*$; $\pi = (P_{абс} - P_0)/P^*$; $\omega = (c - c_0)/c^*$; $\rho_0 = 990$ kg/m³; $\rho^* = 100$ kg/m³; $T_0 = -10°C$; $T^* = 50°C$; $P^* = 120$ MPa; $P_0 = 0.101325$ MPa; $c_0 = 1300$ m/s; $c^* = 500$ m/s.

Equation (9) explicitly expresses the functional dependency between the density of seawater and parameters $T$, $P$, and $c$ and can be useful in marine research. The advantage of this equation is that salinity is not explicitly present in it yet it is anticipated that the density and sonic speed in seawater are functionally related to salinity. An equation like this helps eliminate indirect salinity measurements (via conductivity) by replacing them with direct sonic speed measurements.

Let us analyze the results of measurements taken in the Black Sea and the Sea of Azov.

Based on the two described methods of determining the density of seawater, we calculated two density profiles $p(z)$ depending on the depth $z$ for each of the stations where measurements were taken through vertical probing with the CTD and SVP devices. Figures 3 and 4 show density profiles for the stations in the Black Sea and the Sea of Azov.

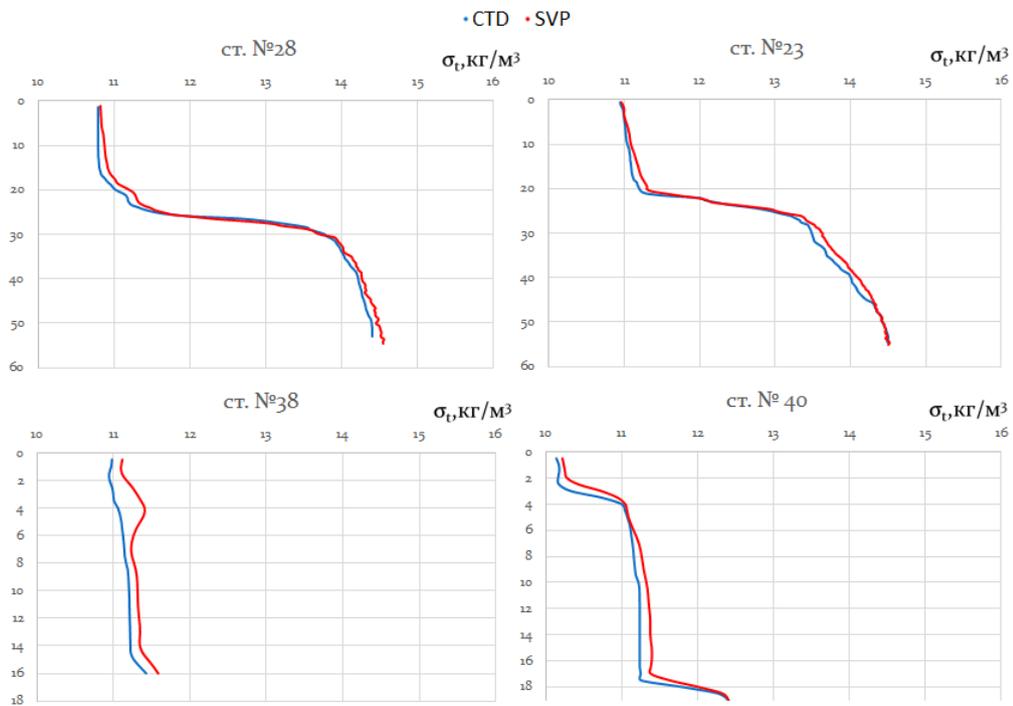

Figure 3. Density profiles at Stations 28, 23, 38, and 40 in the Black Sea

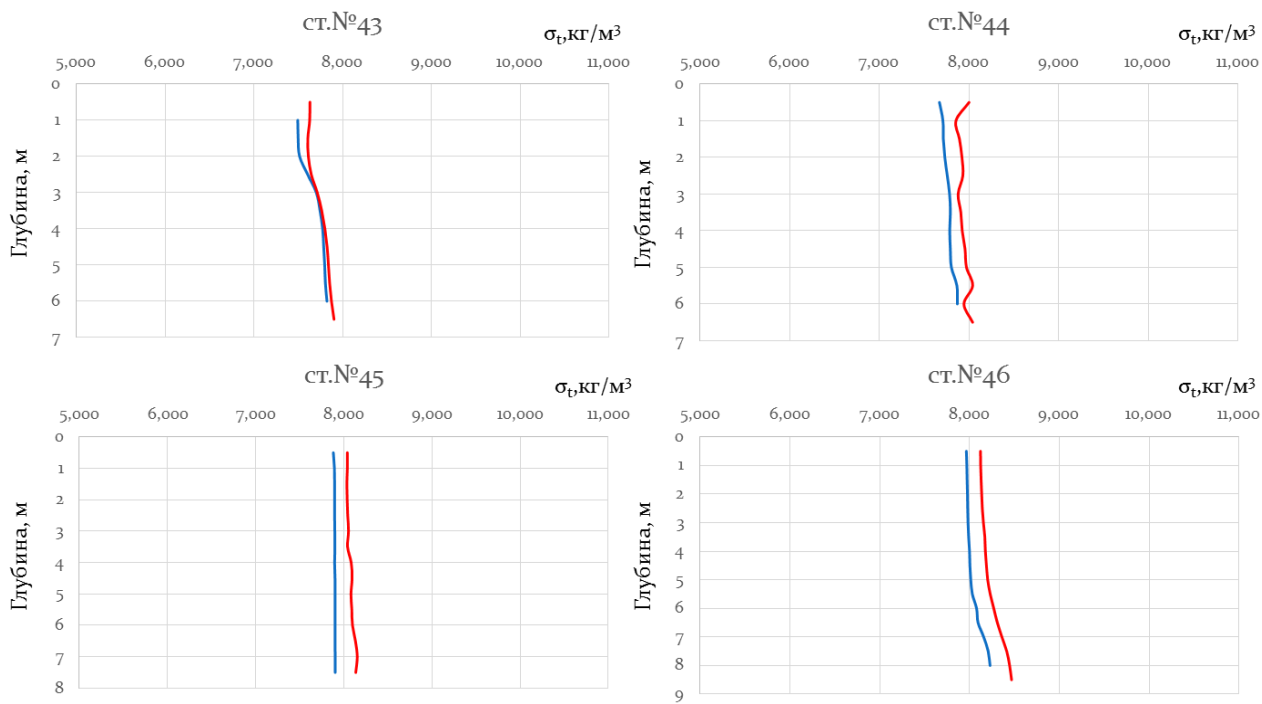

Figure 4. Density profiles at Stations 43-46 the Sea of Azov

These charts show that density profiles have similar shapes, yet the density calculated using $\rho = f(T,P,c)$ exceeds the density calculated using $\rho = f(T,S_A,P)$ on average by $\Delta\rho = 0.20$ $kg/m^3$.

**Difference analysis of vertical density profiles.**

Using the measurement data obtained by the CTD and SVP devices, we performed a difference analysis of vertical density profiles for the stations in the Black Sea and the Sea of Azov. For each of the stations, we found the difference (delta) of densities at various horizons at a depth increment of 1 m. The total average density delta for the Black Sea was 0.23 $kg/m^3$ with the root-mean-square deviation of 0.05 $kg/m^3$. The total average density delta for the Sea of Azov was 0.16 $kg/m^3$ with the root-mean-square deviation of 0.03 $kg/m^3$. The distribution of density delta across the density horizons is shown in Figure 5. For the Black Sea, $\Delta\rho$ is marked in blue, and for the Sea of Azov, it is marked in red.

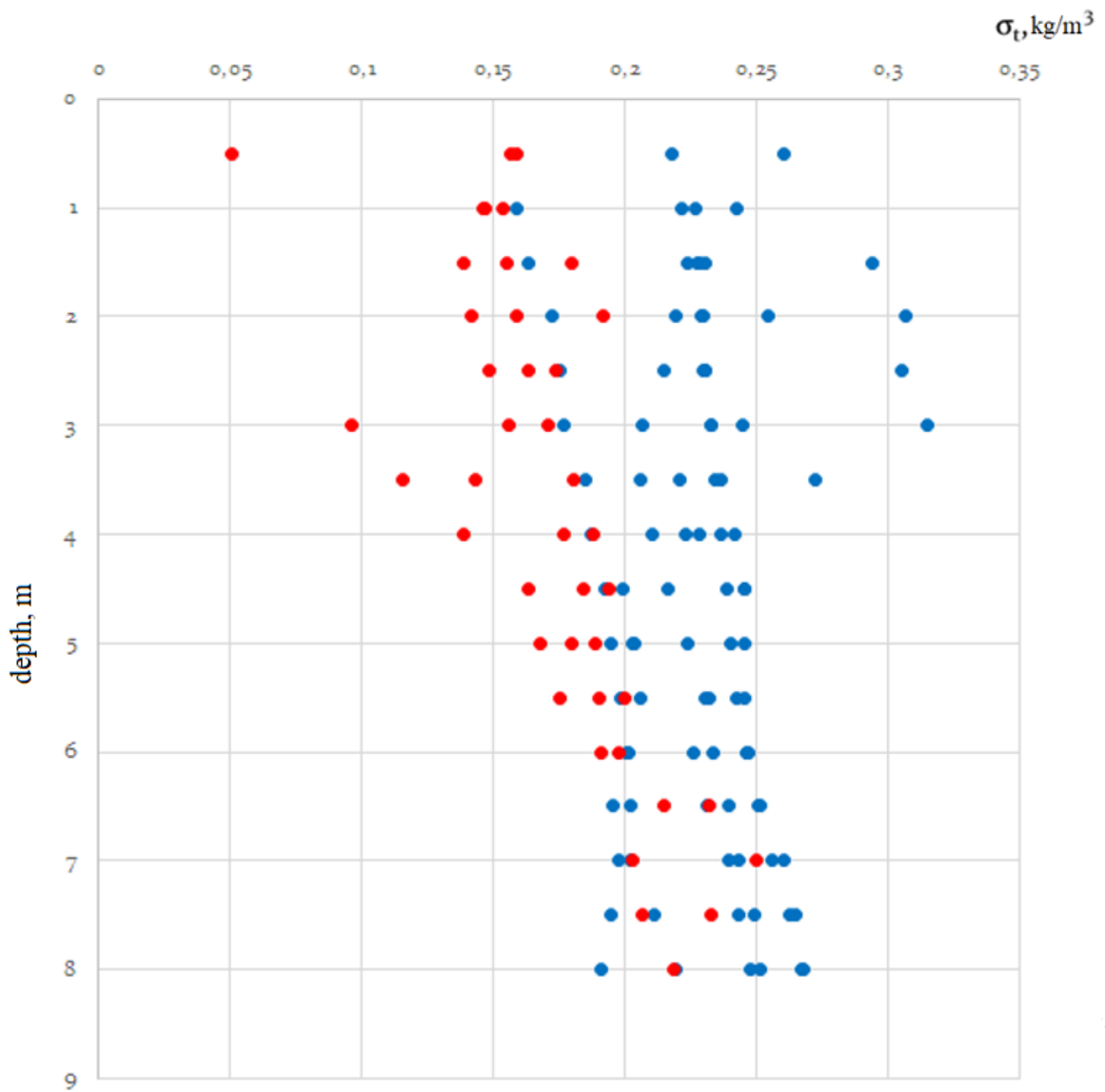

Figure 5. Calculated values for vertical density delta profiles across horizons
according to the measurements taken by the CTD and SVP devices in the Black Sea (blue)
and the Sea of Azov (red)

The use of measurement results obtained at various horizons (2, 4, 6, 8 meters) at Station 38 using the impedance measurement module confirms the possibility of using this parameter to identify suspended materials. Figure 6 shows impedance value charts at various constant frequencies at horizons of 2, 4, 6, and 8 meters. This means, that at 2, 6, and 8 meters, impedance does not change rapidly at different frequencies. However, at about 4 meters and a frequency of 0.8 MHz, the impedance amplitude changes dramatically. The changes in the acoustic signal parameter at the same horizon at the same station also confirm the presence of suspended impurities in seawater.

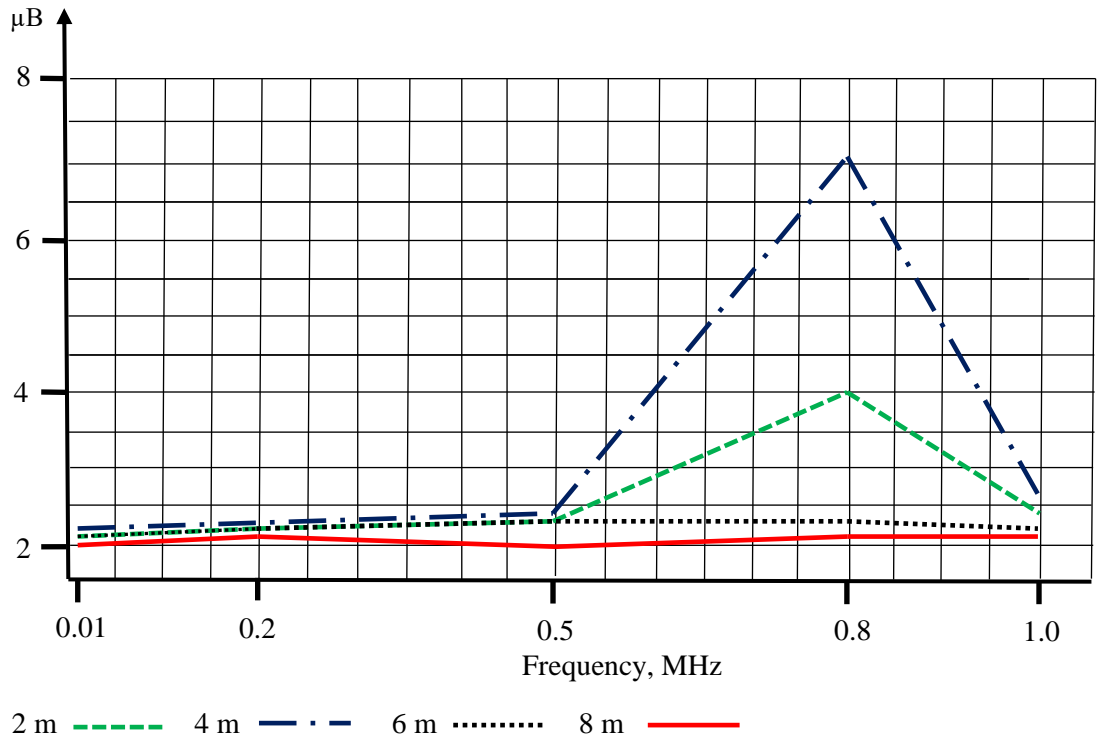

Figure 6. Dependency of impedance meter output voltage amplitude
and frequency at various horizons at Station 38

If we analyze measurements and calculation results shown in Figures 3-6, we can come to the following conclusions.

1. The main instrumental error of the devices is additive and makes an equal contribution to measurement results. Using the same measurement methods, the calculated vertical distributions of density difference $\Delta\rho$ for the Black Sea and the Sea of Azov characterize the difference in the ionic content of the water from these two seas.

2. Some of the stations registered discrepancies from the average density difference $\Delta\rho$ of ±0.1 $kg/m^3$ at specific horizons, which signifies that there are suspended impurities whose impact on electric conductivity is poorly explored and unaccounted for in calculations.

3. At Station 38 (see Figure 3) in the Black Sea, density values obtained through sonic speed measurements by the ISZ device at the horizons of 3-4 m are increased. At the same time, there were almost no changes in electric conductivity at the main operating frequency of the conductivity channel of the CTD probe. The difference analysis of density values shows an anomaly spot of $\Delta\rho$ reaching the maximum value of 0.31 $kg/m^3$.

4. At Station 44 in the Sea of Azov (see Figure 4), there was a drop in density measured through sonic speed. We assume that this was caused by air bubbles, although the electrical conductivity channel did not find such impurities.

5. The use of extra information produced by the acoustic attenuation and impedance metering channel models confirmed the presence of suspended impurities because the measured acoustic attenuation and impedance are directly related to the density and concentration of particles.

Figure 7 shows the vertical profiles of density differences between the measurements taken by the CTD and SVP probes sketched on the map of operations.

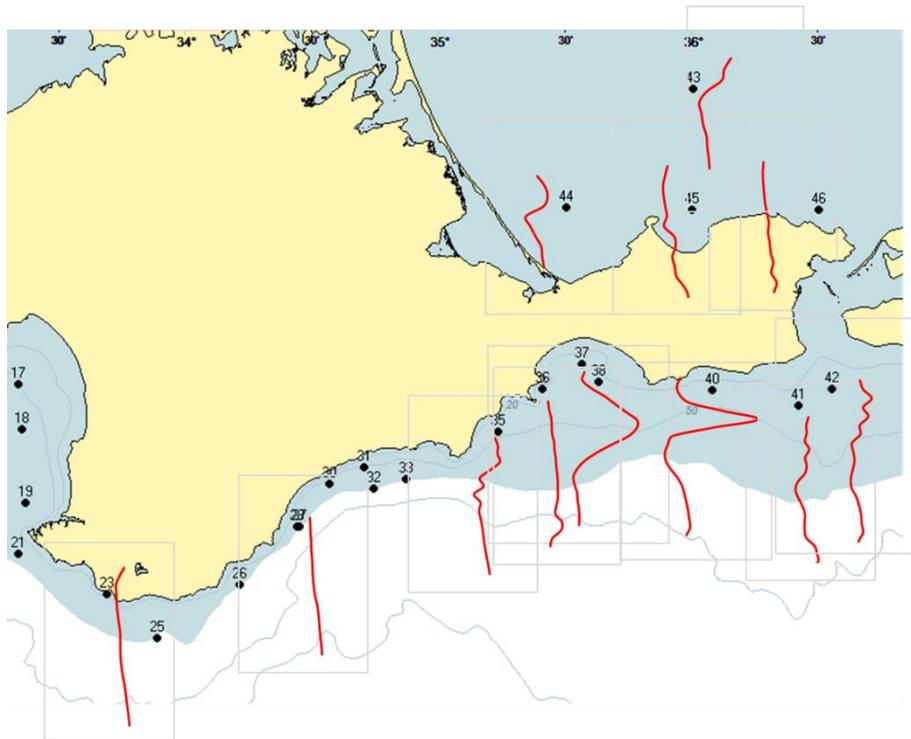

Figure 7. Density difference profiles for CTD and SVP probe measurements sketched on the map of operations

The results of the calculations were used to chart the density delta distribution fields for various horizons. Figure 8 shows the density field charted using the measurements taken by the CTD and SVP probes at a depth of 3 m.

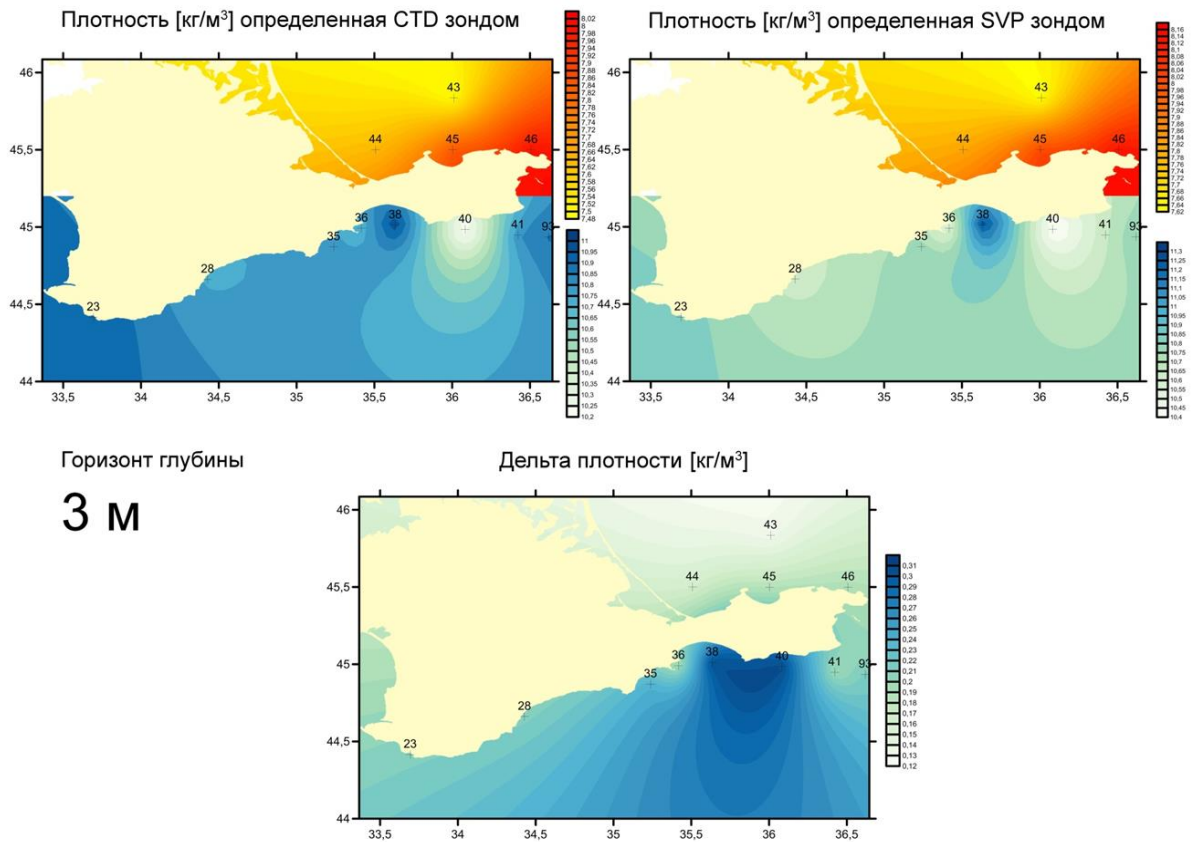

Figure 8. Density field charted using the measurements taken by the CTD and SVP probes at the depth of 3 m during the 96th expedition of Professor Vodyanitsky research ship

The anomaly $\Delta\rho$ reaching a maximum value of 0.31 $kg/m^3$ can be seen clearly in the gulf of Feodosia near the cape Chauda. This is an indirect sign of anthropogenic pollution in the area featuring an oil loading terminal and experiencing regular problems with wastewater.

**Conclusions.**

1. None of the existing methods provides us with a complete understanding of the distribution of density and salinity of seawater, especially in shelf zones, when it has various suspended impurities. Therefore, oceanologic research must combine various instruments based on different physical principles to study the parameters of water. This work suggests improving mass-produced devices without adding new metering channels and using the developed processing methods for the data obtained. This can significantly improve the reliability of measurements and make field research cheaper.

2. The use of the suggested difference technique for metering and processing the results obtained by the CTD and ISZ devices can help assess the correctness of the interpretation of vertical profiles obtained by the CTD probe in the shelf zone when calculating density and salinity. To do this, we need to develop new algorithms and models taking into account the measurements obtained by the ISZ.

3. The developed impedance and acoustic attenuation sensor models allow for the improvement of the functional capabilities of sonic speed and water sensors without structural modification due to the improvement of hardware and software solutions, the improvement of the popular CTD and ISZ devices, and the retrieval of additional qualitative data on the type of suspended impurities in seawater.